\documentclass[aps,noshowpacs,tightenlines,nobalancelastpage,floatfix,twocolumn]{revtex4}
\usepackage{amssymb}
\usepackage{amsmath}
\usepackage{graphicx}
\usepackage{bm}

\begin{document}

\title{Renormalization and small-world model of  fractal quantum repeater networks}
\date{\today }
\author{Zong-Wen Wei$^1$}
\email[Correspondence to : ]{wbravo@mail.ustc.edu.cn}
\author{Bing-Hong Wang$^1$}
\author{ Xiao-Pu Han$^2$}
\affiliation{$^1$Department of Modern Physics, University of
Science and Technology of China, Hefei 230026, China
\\ $2$ Institute of Information Economy and Alibaba Business College, Hangzhou Normal University, Hangzhou 310036, China}

\begin{abstract}
Quantum  networks provide access to exchange of quantum information.
The primary task of quantum networks is to distribute entanglement between remote nodes.
Although quantum repeater protocol enables long distance entanglement distribution, it has been
restricted to one-dimensional linear network.
Here we develop a general framework that allows application of quantum repeater protocol to
arbitrary quantum repeater networks with fractal structure.  Entanglement distribution  across such networks is mapped to renormalization.
Furthermore, we demonstrate that logarithmical times of recursive such renormalization transformations
can trigger fractal to small-world transition, where a scalable quantum small-world network is achieved.
Our result provides new insight into quantum repeater theory  towards realistic construction of large-scale quantum networks.
\end{abstract}
\maketitle

By exploiting the probabilistic prediction nature of quantum mechanics and the nonlocal
 correlation of  entanglement \cite{qe, mpe},
 new technology of quantum  communication  has been developed \cite{qc}.
For example, quantum teleportation  allows faithful
teleportation of unknown quantum states, and quantum cryptography
(Ekert91 protocol)  enables truly secure communication \cite{qc}.
Quantum nodes can store and manipulate photons  locally,
and their interconnection via quantum channel, e.g., optical fiber, gives birth to quantum networks.
Quantum networks are backbone of quantum communication and distributed quantum computing \cite{qi}.
A prototype of quantum network has been reported recently\cite{eqn}.
Quantum networks can be seen as large and complex system of quantum states.
How to characterize and understand such quantum system remains a challenge.
It is no longer  effectively described by a global density operator $\rho$ \cite{qi}.
However, complex networks which describe a wide range of natural and social system \cite{cn, stdy, neta},
provide conceptual basis for in-depth investigation of topological properties of quantum networks.
It involves exciting phenomena. For instance,
entanglement percolation \cite{ep, epqcn, lpep} and the peculiar behavior of quantum random networks \cite{qrn} have received much attention.
What's more, it has opened new perspective for study of entanglements: map complex networks into entangled states,and vice versa \cite{bqsrn, eg}.

As  essential ingredients for quantum information, however, entanglement is such  fragile resource that suffer
fatal photon loss,  decoherence caused by noise, and imperfection of quantum local operations \cite{qc, qr, qrnl, dlcz}. In consequence,
the state fidelity or degree of entanglement decreases exponentially with  channel length.
Quantum repeater protocols (QRP) are
 one of most promising solutions designed to tackle such challenging problem \cite{qc, qr, qrnl, dlcz, qrm, qrpe, qrphe}.
The general principle of QRP  is illustrated  in Fig. 1(a)-(c).
 In principle, QRP allow one to establish long-distance entanglement with fidelity close to unity, while  the required
  time  increases, e.g., polynomially  with channel length (it depends on specific QRP).
Thereby  quantum repeaters hold promise for building large-scale quantum networks.
Then a fundamental problem comes to us: what's the possible topology of quantum repeater network  (QRN),
and how to perform QRP on such network?

In this work, we address this problem, with focus on the interplay between entanglement distribution and topology of quantum networks.
A practical  scenario of entanglement distribution has to consider the complex  topology of quantum networks.
Recent studies  suggest that the topology of quantum networks strongly affects their performance  \cite{ep, epqcn, lpep}.
How well can we say about the topology of QRN?
It's a problem that has not been  seriously considered.
Motivated by the rich and intriguing topology of complex networks, we envisage the topology of QRN as follows.
The exponential decay of fidelity requires that QRN be fractal.
And scale-free network is a plausible option for QRN  \cite{cn,stdy, sf}.
Moreover, as a generic characteristic of real-world networks,  small-world effect \cite{cn, stdy, swn}
is able to reinforce the scalability of quantum networks \cite{lpep}.
We combine these elements with QRN.
Next, so far, QRP have been elucidated  for one-dimensional linear network.
In order to apply QRP to fractal QRN, we draw on concepts and methodology from statistical physics.
 We find a clue to relate implementation of  QRP  to renormalization transformation.
As a result, entanglement distribution over arbitrary fractal QRN
corresponds to a process of successive renormalization transformations.


\section* {Results}
\textbf{Relationship between  QRP and renormalization transformation.}

Here we offer a new perspective on QRP.
As shown in Fig. 1(a)-(c), the operations of QRP exhibit
a hierarchical characteristic and self-similarly nested structure. 
Using standard box-counting method, it's easy to compute fractal dimension $d_{B}=1$.
Historically Kadanoff's seminal picture that spin blocks
  hierarchically nest in a self-similar manner basically
stimulated Wilson's renormalization group theory,
which is renowned as  a powerful tool to the problem of phase transition \cite{re, rg}.
Here we point out that the implementation of QRP with nested structure
actually corresponds to renormalization process in network setting.

As shown in Fig. 1(a)-(c), the quantum channel between
nodes $A$ and $B$ is divided into $N$ segments, and every $\ell_{c}$ consecutive
segments are grouped together into a unit which is extended to larger length-scale.
This procedure is repeated  with $n$ nesting levels, where $N = \ell_{c}^{n}$ \cite{qc, qr, qrnl}.
 If we interpret $\ell_{c}$ as a parameter which is  none other than the  transforming length-scale,
then above procedure can be viewed as   real space renormalization transformation.
Guided by this remarkable idea,
we present an universal  framework  which allows one to apply QRP to arbitrary QRN with fractal structure.

We find clear correspondences between one-dimensional and high dimensional quantum repeaters.
Undoubtedly,  the exponential decay of fidelity imposes strong constraint on the way  nodes are interconnected.
Basically  nodes are  interconnected in accordance with local attachment:
 a node prefers to link to  neighboring nodes via intermediate repeater nodes, rather than distant nodes.
This connection fashion gives rise to fractal structure \cite{mm}.
Then  fractal QRN can be substituted  for the  1D linear network.
And the length of 1D chain is replaced by the diameter of
underneath fractal QRN
\begin{equation}\label{D0}
    D_{0} \sim N^{1/d_{B}},
\end{equation}
namely, largest distance between nodes (in graphic sense, the distance
between two nodes is defined as the number of links along the
shortest path \cite{cn,stdy}).
Compared with segmentation fashion of 1D  chain,
the entire network is divided into boxes, whose size   is equal to $\ell_{c}$,
see Fig. 1(e). It will be desirable if  the nesting level  takes a similar form:
\begin{equation}\label{cnl}
    N^{1/d_{B}} = \ell_{c}^{n_{c}}.
\end{equation}
We will prove that it is indeed the case.
It answers an intuitive problem: how many nesting levels ${n_{c}}$  are required
for quantum  networks  with $N$  nodes?
 Also we will show that it implies a structural transition where small-world is obtained.

\begin{figure}
  \includegraphics[width=8.7cm]{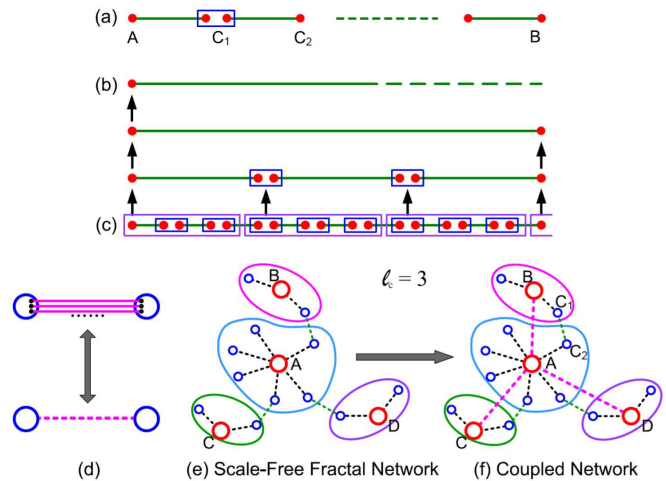}\\
  \caption{\textbf{Renormalization and its relationship with quantum repeaters.}
  Top three panels (also see Ref. \cite{qrnl}): principle of quantum repeaters. (a) The entire channel is divided into $N$ auxiliary segments, avoiding exponential attenuation of fidelity. Entanglements are repeatedly created for each segments. (b) and (c) Nested purification that combines entanglement swapping and purification are successively performed in a hierarchical way.   Adjacent segments are connected and extended  to longer distance, eventually two remote nodes are connected via perfect entanglements.  Bottom panel: schematic illustration of coupled renormalization.
(d) Quantum node  (circle) is a composite system composed of ensembles of qubits (dots). A pair of quantum nodes is connected by multiple copies of bipartite entangled states (solid lines). For simplicity, they are represented by single dashed lines in Fig. 1(e) and (f). (e) Nodes are assigned to different boxes according to the MEMB
algorithm. (f) Single level CR is performed.}
\end{figure}

Renormalization  was successfully
introduced into complex networks by Song \emph{et al.}, uncovering
 the self-similarity  of complex networks \cite{fr}.  A network is renormalized  according to the
box-covering technique \cite{fr} (see  Fig. 1(e)-(f)).
The basic idea  is as follows:
tile the entire network with minimum number of boxes $N_{B}$, where
the distance between nodes within any box is smaller than the
box size, namely, the transforming length-scale $\ell_{B}$. Each box
is then replaced by a supernode.  These supernodes are
connected if there is at least one link between nodes in their respective boxes.
It defines the fractal dimension
$d_{B}$ in terms of a power law:
\begin{equation}\label{fD}
    \frac{N_{B}}{N} \sim \ell_{B}^{-d_{B}}.
\end{equation}
Apply this transformation $R_{\ell_{B}}$ to a fractal network $G_{0}$, which is scale invariant,
then we have $R_{\ell_{B}}(G_{0})=G_{0}$.

    Several algorithms \cite{fd1, fd} have been proposed to coarse-grain complex
networks, nevertheless, not all of them are useful here.
We choose  the MEMB algorithm to divide the entire network into boxes  \cite{fd}.
It's a geometric algorithm which has the advantages of
guaranteeing connectivity within boxes, isolating hubs of
different boxes and avoiding overlap between boxes.

 We then introduce some modifications.  For clarity, this is  rephrased in network language.   We select the hub of a
box (most connected node) as  representative node, playing the role of
 supernode.  If there is  one link between two  boxes, then add one link connecting their hubs.
Otherwise, no  links are attached. Above presentation
instructs us to create shortcuts (long-distance entangled state with high fidelity) between which pair of representative nodes.
 As shown in Fig. 1(f), the corresponding shortcuts denoted by pink links
 form a coarse-grained network (CGN). As a result, the CGN is reconstructed and coupled to the
initial network. Because of the self-similarity of underling network, the resulting CGN is
topologically equivalent to the original one, and above renormalization processing
 can be iteratively applied to previous CGN with fixed $\ell_{c}$,
  and so on until the critical nesting level (see Eq. (\ref{NL})).
  Notice that the requirement of local attachment is  fulfilled throughout entire process.
Eventually, the superposition of each level CGN forms a new network, namely, the coupled network (CN).
For the sake of distinction between the standard
 and this modified renormalization, we name the later coupled renormalization (CR).

We have generalize quantum repeaters to arbitrary high but finite fractal dimension,
 in the sense that when $d_{B}=1$, CR is automatically reduced to 1D quantum repeaters.
  One possible application of CR for 1D linear network is shown in Fig. 1(c).
In 1D case,  each representative node  is merely connected to  two representative nodes (nodes below arrows).
 In regard to fractal QRN, however, a representative node (red circles) has
 probability $P(k')$  to  link $k'$  representative nodes (it depends on the degree distribution) via
 the  short path composed of   entangled links  between two boxes (green links).
 For instance, in Fig. 1(f), hubs $A$ and $B$ are connected through a path $B$-$C_{1}$-$C_{2}$-$A$, whose length is $\ell_{c}=3$.
 Each of such a path corresponds to one unit of  1D case (segments between two arrows in Fig. 1(c)).
 Then  shortcuts  are established between representative nodes,
with quantum operations  identical to 1D case. In other words, its  definite  physical realization relies on which QRP is utilized.
 Entanglement swapping \cite{es} and purification \cite{epu, epuc, eput} are two most important quantum operations.
 By entanglement swapping, adjacent entangled links are connected and extended to two representative nodes.
 To obtain high fidelity entangled states, entanglement purification is required, which extracts nearly
 perfect entangled states from states with lower degree of entanglement.
 Alternatively, some QRP use quantum error correction \cite{qrm, qrpe, qrphe}. This class of QRP  could circumvent
the  probabilistic fashion of purification-based protocols, and relax the strong requirement of long-lived quantum memory \cite{qmsn, qsep, hsbe}.
Thus they have potential  to extend entanglement to longer distance and speed up communication rate.

\textbf{Why  small-world and scale-free properties are relevant for quantum networks.}

 Despite  the great diversity of real-world networks, they share some common features.
 Most of real-world networks are scale-free networks with small-world property \cite{cn, stdy}. A network is scale-free if
the probability to find a node with $k$ links $P(k)$, follows a power law $P(k)\sim k^{-\gamma}$ (for real-world networks, $2<\gamma <3$).
This is quite different from Poisson distribution for Erd\H{o}s-R\'{e}nyi random graphs  \cite{cn,stdy, sf}.
Small-world  is an influential  concept describing such a phenomenon:
despite the large size of networks, on average,
  any two nodes are separated by relatively short distance, which typically scales as:
$\bar{\ell} \sim \ln N$ \cite{cn,stdy, swn}.
These features  play a dominant role on  dynamic functions of complex networks.
One naturally wonders whether the two fundamental characteristics make a difference to quantum networks

The realistic  significance can be illustrated from the perspective of limited-path-length entanglement percolation \cite{lpep}.
Without quantum processing such as entanglement purification, the fidelity of communication along   a long noisy path of imperfect entangled states decreases exponentially \cite{qrnl, lpep}. Thus it yields a very short distance $\ell$ of faithful  communication, which severely limits effective size of quantum networks $N_{e}$.
Things seem to be bad. However, if $\bar{\ell}<\ell$,  it turns out that most of nodes are reachable  within the distance of reliable communication.
For $d$-dimensional regular lattice, $\bar{\ell}\sim N^{1/d}$,
and we have $N_{e} \sim \ell^{d}$. In contrast, $N_{e} \sim e^{\ell}$ for small-world networks.
We therefore need a quantum small-world network,
so that  without further quantum processing, this desirable topological effect alone suffices to effectively
mitigate the limitation of noise, and enhance the  scalability  of quantum networks.

\begin{figure}
  \includegraphics[width=8.7cm]{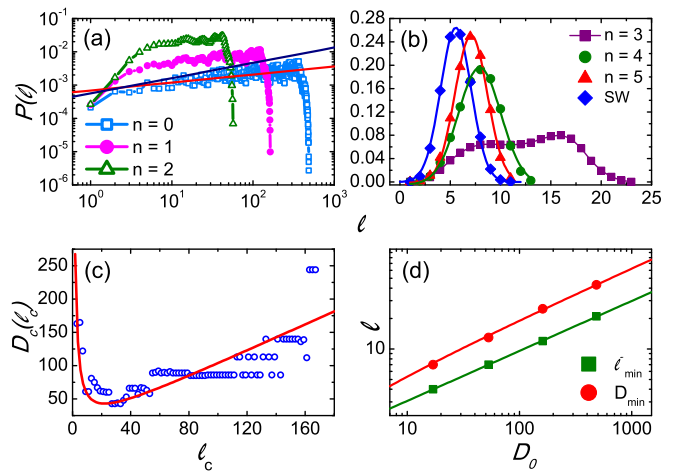}\\
  \caption{\textbf{ Statistical properties of CN.}
  In this example, $t=5$, $N=9375$, $\ell_{c}=3$.
  (a) and (b) Distance distribution of  CN with different nesting levels. (a) is log-log plot of $P(\ell)$ versus $\ell$,
   the slope of the upper line is $d_{B}-1\approx0.46$ (analytical estimate), and the lower line 0.24 (fitting).
    (c) Prediction of diameter of single  level CN as a function of $\ell_{c}$. (d) Log-log plot of the minimal  and average diameter of different size of single level CN.}
\end{figure}

The  striking  effect of small-world originates from the existence of shortcuts between remote nodes.
A quantum network without shortcuts is not  small-world network.
Consequently, it's  rather difficult to extend the influential concept of small-world to large-scale quantum  networks, where shortcuts  are not directly available.
 CR can generate a set of shortcuts, which eventually leads to small-world transition at a relatively small nesting level.
 We will clarify it in subsequent section.
While  CR can be applied to  fractal networks such as
Erd\H{o}s-R\'{e}nyi random graphs  at criticality \cite{fr},
apparently, such topology is not a realistic option.
We propose that QRN are scale-free and fractal  observed at arbitrary spacial scale.
We justify this proposal with  twofold  reasons.



Entanglement  percolation is a  combination of entanglement swapping and classical entanglement percolation\cite{ep}.
In their framework, the threshold  actually depends on the final topology.
Notably, the percolation  threshold for scale-free networks can be vanishing \cite{rIn, rprl, rpre}.
It means that, as far as scale-free  networks, classical entanglement percolation  is such a  good  strategy
that the critical  amount of entanglement required for the presence of giant cluster is zero in asymptotic limit.
Therefore, scale-free QRN has exceptional ability to  preserve connectivity in the presence of noise.
In addition, it's strongly supported  by three facts.
 Other than local operations, classical
communication is indispensable between quantum nodes.
And entangled photon pairs can be transmitted
through commercial telecom fiber.
Thus, to some extent, QRN is embedded in  classical communication networks.
Whereas both phone call networks and Internet are  scale-free networks \cite{cn, stdy}.
If QRN is scale-free, we can make full use of the existing network infrastructures,
without significantly altering them.
Above facts suggest that scale-free network is a plausible  and eligible candidate for QRN.

Without loss of generality, we apply CR  to a scale-free fractal network generated by the minimal model (see Methods).
  Let $G_{n}$ be the $n$th level  CGN. According to renormalization group theory,
 $G_{n}=R_{\ell_{c}}(G_{n-1})=R_{\ell_{c}}^{n}(G_{0})=R_{\ell_{c}^{n}}(G_{0})$.
 So $G_{n}$ can be seen as larger-scale network enlarged from $G_{n-1}$,
or equivalently it arises from single level CR with  transforming length-scale $\ell_{c}^{n}$.
Larger-scale CGN here collectively  act as shortcuts of the underlying smaller-scale ones,
which drastically change the topology in such a way that  nodes
are globally separated by short path of entangled links.
This can be further unveiled by   the squeezed distance distribution  which follows Gaussian distribution (see Fig. 2(b)).
Hence, in the end, a hierarchically nested quantum small-world network  is produced.

\textbf{Proof of fractal to small-world transition.} We proceed to make
 it clear  whether single or iterative CR will lead to fractal to small-world transition.
   Two analytical proofs with numerical simulations are provided.
A rigorous and reliable method is to observe the behavior of average
degree under renormalization flow  (see Methods) \cite{sw}.
  In regard to single level CR, the expected transition does not arise.
  However,  it's safe to say that iterative CR can give rise to  fractal
to small-world transition. Evidences
for the transition  displayed in  Fig. 3  conform above conclusion.
Nonetheless, the average diameter of CN, namely, the signature
 of small-world  is unclear.

We begin with analyzing the impact
of single level CR, and then generalize it to iterative  case.
Plugging Eq. (\ref{fD}) into Eq. (\ref{D0}),  we immediately obtain the diameter of CGN
$D_{B}(\ell_{c}) \sim D_{0}/\ell_{c}$. In order to compute $D_{C}(\ell_{c})$, the
diameter of CN, we suggest a hierarchical
routing method which exploits the hierarchical structure and convert it into a routing problem.  A path  connecting two remote nodes
is divided into two parts: one part links one of the nodes and its corresponding hub
within the box, and the other part, consisting of only representative nodes,
connects the two hubs.
 The total length of the first part is
approximately $\ell_{c}-1$, and that of the second is $D_{B}(\ell_{c})$. We thus have
\begin{equation}\label{Dc}
D_{C}(\ell_{c}) \approx \frac{D_{0}}{\ell_{c}}  +\ell_{c}-1,
\end{equation}
see Fig. 2(c). Take note that   there is an optimal transforming length-scale, $\ell_{o}=\sqrt{D_{0}}$,
 which  yields minimal diameter $D_{min}(N)=2\sqrt{D_{0}}-1$.
Meanwhile, numerical simulation gives the corresponding minimal average diameter $\bar{\ell}_{min}(N)=D_{min}(N)/2\approx\sqrt{D_{0}}$ (see Fig. 2(d)).
An analytical approximation shows that $\bar{\ell}_{min}(N)$ scales as a power-law too (see  supplemental information).
The behavior of $\bar{\ell}_{min}$ suggests that single level CR is unable to trigger the transition, which is
 consistent with above conclusion.

\begin{figure}
  \includegraphics[width=8.7cm]{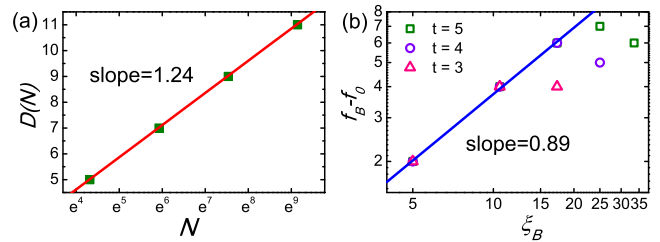}\\
  \caption{\textbf{ Evidences for QRN with small-world property.}
  (a) Log-log plot of Eq. (\ref{cd})  with parameters $\ell_{c}=3$, and $d_{B}\approx1.46$.
  The analytical prediction (straight line) matches well with numerical simulation (square).
  (b) Log-log plot of $f_{B}-f_{0}$ versus $\xi_{B}$.}
\end{figure}

 Now let's consider multi-level CR with fixed box size. In analogy with above results, it's easy to
obtain the diameter of iterative CN $D(n,\ell_{c})$ for small $\ell_{c}$, using Eq.
 (\ref{Dc}) with recursive derivation, we find
\begin{equation}\label{iD}
    D(n,\ell_{c}) \approx \frac{D_{0}}{\ell_{c}^{n}}+n(\ell_{c}-1),
\end{equation}
where $n$ is  nesting level of CR.   Taking into account  finite size effect,  Eq. (\ref{iD})  holds on condition that $\ell_{c}\ll \ell_{o}$.

Remarkably,  the first term decays exponentially,  whereas
the second increases linearly.
 Hence, $D(n,\ell_{c})$ is governed by the linear term and grows slowly.  We readily obtain the criterion for the transition:
 \begin{equation}\label{cond}
    \ell_{c}^{n_{c}} \sim D_{0},
\end{equation}
implying that $D(n_{c},\ell_{c}) \approx n_{c}(\ell_{c}-1)$ for large
$N$. It's desirable that both $n_{c}$ and $D(n_{c},\ell_{c})$
 increases logarithmically
 with size of network, since
\begin{equation}\label{cd}
  D(n_{c},\ell_{c})
\approx \frac{\ell_{c}-1}{d_{B}\ln \ell_{c}}\ln N,
\end{equation}
and
\begin{equation}\label{NL}
n_{c}\sim\frac{\ln N}{d_{B}\ln \ell_{c}}.
\end{equation}
How to appreciate the implication of $n_{c}$ now is evident. We identify $n_{c}$ as critical nesting level,
at which small-world is achieved. Direct evidence is  shown by Eq. (\ref{cd}).
What's more, it's  exactly in agreement with simulation result, see Fig. 3(a).
Here $n_{c}$ is inversely proportional to fractal dimension $d_{B}$,
 indicating its topological interdependence.   When $d_{B}=1$, Eq. (\ref{NL}) reproduces
the result of 1D case. While  $d_{B}\rightarrow\infty$,  $n_{c}\rightarrow 0$,
means that the initial network is already small-world, perfectly consistent  with conclusion that
  when  $d_{B}\rightarrow\infty$, these networks are small-world without fractality \cite{mm, trf}.

\section* {Discussion}

The highlights of our scenario are as follows. Our scenario is a fairly general framework. CR is compatible with
various QRP based on the aforementioned principle.
  In principle, CR is  applicable
to arbitrary quantum networks with fractal structure, not restricted to the example of
scale-free networks. And the transition will arise as long as  Eq. (\ref{cond}) or  Eq. (\ref{NL}) is satisfied.
It's not difficult to check that the unique requirement of the whole derivations is the fractality of QRN.
Furthermore, the collective distribution of shortcuts across  entire network  is mapped
to renormalization transformation, where the self-similar fashion
of operations is preserved at  network level, while the
scale-free fractal structure is kept at all length-scales.
 Each level transformation enlarges underling network into larger-scale  QRN.
 The simultaneous logarithmical scaling of critical nesting level and corresponding diameter suggests
 that to achieve small-world, CR is operable even for   QRN of large size.
Moreover, thanks to the scale-free nature, CN is particularly resilient to random failure of quantum nodes.


To summarize, we have generalized one-dimensional quantum repeaters to high fractal dimension by introducing an approach called CR,
which relates entanglement distribution over arbitrary fractal QRN  to recursive renormalization transformations.
We assume that QRN is fractal and scale-free, which is the case for a large number of real-world networks.
In spite of the large size of QRN, there exists a relatively small critical nesting level for CR, at which small-world is obtained.
Small-world seems to be a necessary element for a scalable quantum network.
Our study suggests that concepts and tools from statistical physics will play an important role in the joint study.
 It has conceived another significant direction which  may open new avenue to address the outstanding issue of complexity.
That is, quantum simulation of dynamic process
on complex networks  or design of quantum algorithms for sophisticated
questions in network science   \cite{qpr, gn, qnr}.
All of these attempts may dramatically alter the landscape of both fields.

\section* {Methods}
\textbf{Minimal model for Scale-Free networks.}

    The growth of the network is actually the inverse
procedure of renormalization \cite{mm}.
 we begin from a triangle at time step
$t=0$.

 (i) At time step $t+1$, $m$ new nodes are connected to the
endpoints of each link $l$ generated at time step $t$.

(ii) Then,
with probability $1-e$, we remove  link $l$ of time step $t$,
and add one new link connecting a new pair of  nodes attached to the endpoints of link
$l$.

(iii) Repeat (i) and (ii) recursively until
the wanted time step.

This model produces a scale-free network with
degree distribution exponent $\gamma=1+\ln (2m+1)/\ln (m+e)$. We are
particularly interested in two distinct types of networks with $e=0$
or $e=1$, where a pure fractal network with fractal dimension
$d_{B}=\ln (2m+1)/\ln 3$, and a nonfractal, small-world network (SW) are achieved
respectively. For simplicity and
 without loss of university, here let $m=2, e=0$, then $d_{B} \approx 1.46$.

\textbf{Renormalization group method for proof of fractal to small-world transition.}


Literature \cite{sw} studied networks
constructed by randomly adding links
to a fractal network with probability $p(\ell)\sim \ell^{-\alpha}$.
Let $f_{0}$,  $f^{'}$ and $f_{B}$ be
the average degree of the initial, the new and renormalized network  respectively. Then
\begin{equation}\label{rf}
    f_{B}-f_{0}=(f^{'}-f_{0})\xi_{B}^{\lambda},
\end{equation}
 where $\xi_{B}=\ell_{B}^{d_{B}}$. Let $s=\alpha/d_{B}$.  If $s>1$, $\lambda=2-s$, otherwise $\lambda=1$.  Notice that when $s=2$, $\lambda=0$,
 a stable phase corresponding to fractal network is separated
from the unstable phase moving toward complete graph, where
small-world  is achieved.

With  purpose of finding the
location of CN in the phase diagram, we have to calculate the exponent $\lambda$.
 In our model, on one hand,
 \begin{equation}\label{pl}
    p(\ell)\sim \ell^{-d_{B}}/\ell^{d_{B}-1}=\ell^{-(2d_{B}-1)},
 \end{equation}
  so $\alpha\approx2d_{B}-1$, and $s=2-1/d_{B}$, $\lambda=2-s=1/d_{B}\approx 0.68$.
 On the other hand, numerical simulation shows that
 $\lambda\approx 0.89$ (see Fig. 3(b)). The apparent deviation is mainly caused by the rough approximation that $p(\ell)$  follows a power law
(see Fig. 2(a)).
 In spite of the deviation, it's definite that $\lambda\gg0$. Thus, CN belongs to the unstable
phase and  multi-level CR can give rise to  fractal
to small-world transition.

For single level CR, we have
\begin{equation}\label{scr}
    f_{B}-f_{0}=2\ell_{c}^{-d_{B}}N/N_{B}=2(\ell_{B}/\ell_{c})^{d_{B}}.
\end{equation}
When $\ell_{B}>\ell_{c}$, the links we added will not emerge in the renormalized network  with length-scale $\ell_{B}$  \cite{sw}.
We only need to consider one case $\ell_{c}=\ell_{o}$, which diverges in the large size limit. So
$f_{B}-f_{0}\rightarrow0$, and $\lambda\ll0$.   Thus, coinciding with prediction of
equation $\bar{\ell}_{min}(N)=\sqrt{D_{0}}=N^{1/d_{B}}$,  single level CN stays in the stable phase, the expected transition does not occur.

\section* {Acknowledgements}
We thank Chaoming Song and Liang Jiang for helpful  suggestions. 
This work is funded by the National Natural Science Foundation of China (Grant Nos. 11275186, 10975126, 91024026, 11205040),  the Major Important Project Fund for Anhui University Nature Science Research (Grant No. KJ2011ZD07) and the Specialized Research Fund for the Doctoral Program of Higher Education of China (Grant No. 20093402110032).

\section* {Author contributions}

Z.-W.W. conceived the research, performed analytical and numerical calculations, and wrote the manuscript.
X.-P.H. discussed the manuscript and prepared the figures. B.-H.W. supervised the research.

\section* {Additional information}
\textbf{Competing financial interests:} The authors declare no competing financial interests.

\vspace*{40pt}

\section* {Supplemental Information}
  
\textbf{Calculation of average diameter for single CR.}
\begin{figure}
  \includegraphics[width=8.7cm]{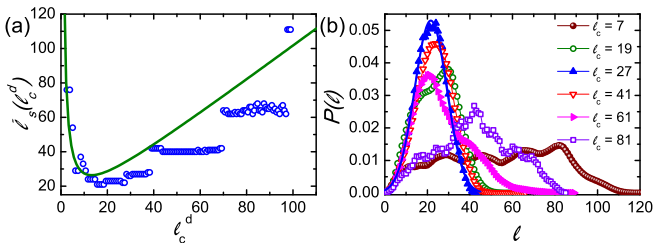}\\
  \caption{  (a) Prediction of average diameter of single CN.
  (b) Distance distribution of single CN for different transforming length-scales.}
\end{figure} 

For scale-free fractal networks which are not small-world,
the average diameter $\bar{\ell}$ \cite{ad} and diameter $D_{0}$ are as follows:
\begin{equation}\label{aD}
    \bar{\ell} \sim N^{\gamma-2/\gamma-1},
\end{equation}
\begin{equation}\label{D0}
D_{0} \sim  N^{1/d_{B}},
\end{equation}
After renormalization, the size of the new network
  $ N_{B} \sim N\ell_{B}^{-d_{B}}$.
And the degree of a node  $k^{'}$ is related with
its original degree $k$ by a power law: $k^{'} \sim
\ell_{B}^{-d_{k}}k$, where $\gamma=1+d_{B}/d_{k}$ \cite{fr}.
 Combining Eqs. (\ref{aD}) and
(\ref{D0}) gives $\bar{\ell} \sim D_{0}^{d}$, where $d =
d_{B}-d_{k}$. So the average diameter of the renormalized network is
$(D_{0}/\ell_{c})^{d}$, and within a box of size $\ell_{c}$, it's $\ell_{c}^{d}$.
 Accordingly, the  average diameter for single CN   approximately behaves as follows:
 \begin{equation}\label{ub}
    \bar{\ell}_{s}(\ell_{c})\approx (\frac{D_{0}}{\ell_{c}})^{d} +\ell_{c}^{d}.
 \end{equation}
Once again,  the optimal transforming length-scale $\ell_{o}=\sqrt{D_{0}}$,
 which yields minimal  average diameter

  \begin{equation}\label{lm}
    \bar{\ell}_{min}(N)=2D_{0}^{0.5d},
  \end{equation}
  see  Fig. 4(a). 
And it is in good agrement with numerical results.
 In fact, numerical simulation gives
 $\bar{\ell}_{min}(N)=D_{min}(N)/2\approx\sqrt{D_{0}}$. These
can be further demonstrated  distance distribution between nodes. As shown in Fig. 4(b), in particular,
with respect to the optimal  transforming length-scales $\ell_{o}=27$, it shrinks into Gaussian distribution.
Naturally, it's not surprising that $\bar{\ell}_{min}(N)=D_{min}(N)/2$.
Although Eq. (\ref{lm}) is just an approximation of $\bar{\ell}_{min}(N)$,
it successfully predicts that single CR is unable to trigger  fractal to small-world transition.

.

\end{document}